\documentclass{webofc}
\usepackage[varg]{txfonts}   
\begin{document}
\title{Temperature dependence of the properties of open heavy-flavor mesons}
%
%

\author{\firstname{Glòria} \lastname{Montaña}\inst{1}\fnsep\thanks{\email{gmontana@fqa.ub.edu}} \and
        \firstname{Àngels} \lastname{Ramos}\inst{1} \and
        \firstname{Laura} \lastname{Tolos}\inst{2,3,4,5} \and
        \firstname{Juan M.} \lastname{Torres-Rincon}\inst{6}
}

\institute{Departament de F\'isica Qu\`antica i Astrof\'isica and Institut de Ci\`encies del        Cosmos (ICCUB), Facultat de F\'isica,  Universitat de Barcelona, Mart\'i i Franqu\`es 1, 08028 Barcelona, Spain
\and
    Institute of Space Sciences (ICE, CSIC), Campus UAB, Carrer de Can Magrans, 08193, Barcelona, Spain
\and
    Institut d'Estudis Espacials de Catalunya (IEEC), 08034 Barcelona, Spain
\and
    Faculty of Science and Technology, University of Stavanger, 4036 Stavanger, Norway
\and
    Frankfurt Institute for Advanced Studies, Ruth-Moufang-Str. 1, 60438 Frankfurt am Main, Germany
\and
    Institut f\"ur Theoretische Physik, Goethe Universit\"at Frankfurt, Max-von-Laue-Str. 1, 60438 Frankfurt, Germany
          }

\abstract{%
  We address the modification of open heavy-flavor mesons in a hot medium of light mesons within an effective theory approach consistent with chiral and heavy-quark spin-flavor symmetries and the use of the imaginary-time formalism to introduce the non-zero temperature effects to the theory. The unitarized scattering amplitudes, the ground-state self-energies and the corresponding spectral functions are calculated self-consistently. 
  We use the thermal ground-state spectral functions obtained with this methodology to further calculate 1) open-charm meson Euclidean correlators, and 2) off-shell transport coefficients in the hadronic phase. 
}
\maketitle
%
Heavy quarks, produced in large abundance in the initial stage in heavy-ion collisions at LHC and RHIC, are a unique probe of the matter formed. Open heavy-flavor mesons are created at the confinement transition and they interact with the light mesons in a hot medium at vanishing baryon density. In this paper we study open heavy-flavor mesons in hot matter, while comparing our results to lattice QCD (LQCD) data and analyzing the implications for their propagation in heavy ion collisions.

\section{Effective theory at finite temperature}
\label{sec-1}
The scattering of open heavy-flavor ($D=\{D,D_s\}$) mesons off light mesons ($\Phi=\{\pi, K, \bar{K}, \eta\}$) in a mesonic thermal bath is described in \cite{Montana:2020lfi,Montana:2020vjg} with an effective Lagrangian consistent with chiral and heavy-quark spin-flavor symmetries. The $s$-wave scattering amplitudes are unitarized by solving the Bethe-Salpeter equation in coupled channels, $T_{ij}=V_{ij}+V_{ik}G_kT_{kj}$,
where $T$ is the unitarized amplitude, $V$ is the interaction kernel, $G$ is the heavy-light meson-meson ($D\Phi$) two-body propagator, and the subindices $i,j,k$ refer to a particular $D\Phi$ channel. The thermal corrections are implemented in this model by employing the imaginary-time formalism and dressing the heavy meson in the loop with its spectral function,
\newpage
\begin{equation}\label{eq:loop}
    G_{D\Phi}(E,\vec{p};T)=\int\frac{d^3q}{(2\pi)^3}\int d\omega\int d\omega'\frac{S_{D}(\omega,\vec{q};T)S_{\Phi}(\omega',\vec{p}-\vec{q};T)}{E-\omega-\omega'+i\varepsilon}[1+f_\omega+f_{\omega'}] \ ,
\end{equation}
where the momentum integral is regularized with a cut-off and $f_E\equiv f(E,T)$ is the Bose-Einstein distribution function. A $\delta$-type spectral function is considered for the light meson, as the modification of the pion with temperature is rather moderate and the effect on the heavy mesons is even smaller.
The spectral function of the heavy meson is given by
\begin{equation}\label{eq:S}
    S_{D}(\omega,\vec{q};T)=-\frac{1}{\pi}{\rm Im\,}\mathcal{D}_{D}(\omega,\vec{q};T)=-\frac{1}{\pi}{\rm Im\,}\Bigg(\frac{1}{\omega^2-\vec{q}^2-m_{D}^2-\Pi_{D}(\omega,\vec{q};T)}\Bigg) \ ,
\end{equation}
where the self-energy is obtained from closing the light-meson line (i.e. the pion line for a pionic bath) in the corresponding $T$-matrix element,
\begin{equation}\label{eq:pi}
    \Pi_D(E,\vec{p};T)=\frac{1}{\pi}\int\frac{d^3q}{(2\pi)^3}\int d\Omega\frac{E}{\omega_\Phi}\frac{f(\Omega,T)-f(\omega_\Phi,T)}{E\,^2-(\omega_\Phi-\Omega)^2+i\varepsilon}{\rm Im\,}T_{D\Phi}(\Omega,\vec{p}+\vec{q};T) \ .
\end{equation}
Self consistency is achieved by solving iteratively Eqs.~(\ref{eq:loop}) to (\ref{eq:pi}). 

\begin{figure}[htbp!]
\centering
\includegraphics[width=10cm,clip]{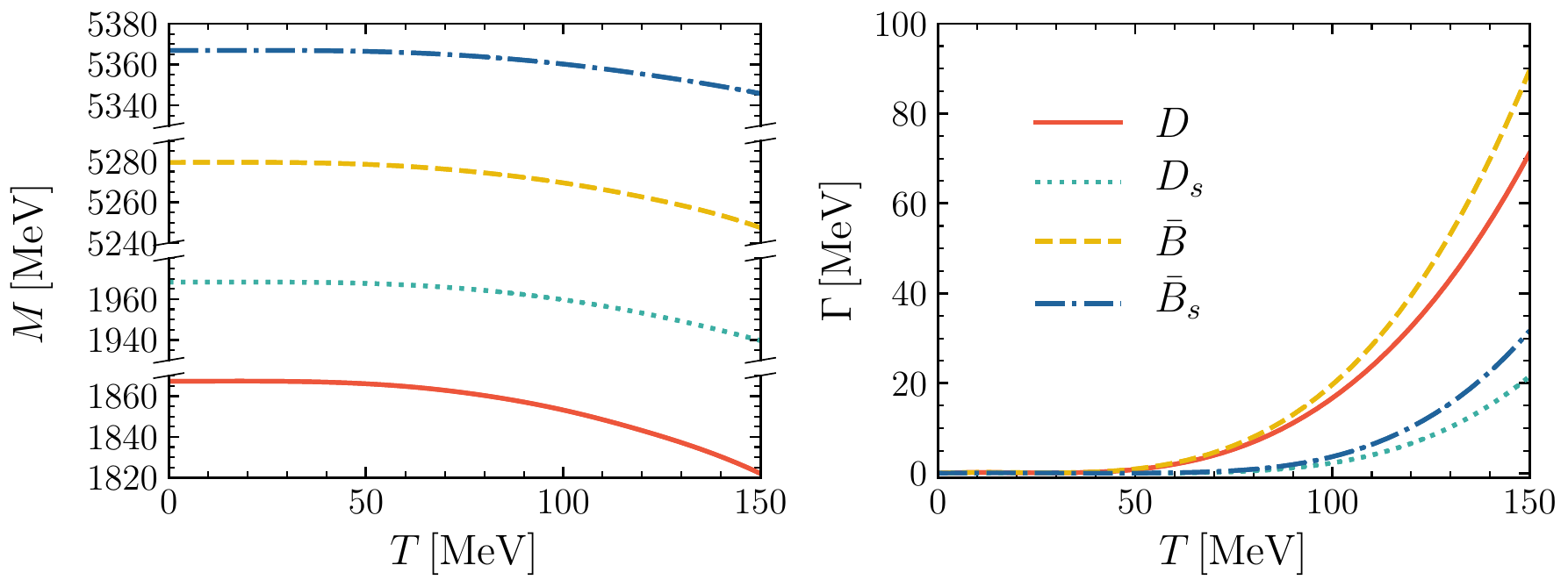}\vspace{-0.2cm}
\caption{Thermal modification of the mass (left panel) and the width (right panel) of the open heavy-flavor ground states.  (Partially taken from \cite{Montana:2020vjg}.)}
\label{fig:M_Width_gs}       
\end{figure}

The properties of the heavy mesons and their thermal modification are contained in their spectral functions. The temperature evolution of the masses and the widths of the $D$ and $D_s$ mesons in a pionic bath below the deconfining transition temperature $T_c$ is shown in Fig.~\ref{fig:M_Width_gs}. The results for the open-charm vector counterparts are found in \cite{Montana:2020vjg}. The heavy-quark spin-flavor symmetry allows us to extend the model to the open-bottom sector upon the replacement $D\rightarrow\bar{B}$. The thermal effects on the masses and widths of the $\bar{B}$ and $\bar{B}_s$ mesons are also displayed in Fig.~\ref{fig:M_Width_gs}, being of the same magnitude of those on the open-charm states. We observe a decrease of the masses of some tens of MeV at $T=150$ MeV, as well as a widening of the same order.

\begin{figure}[bt!]
\centering
\includegraphics[width=0.95\textwidth,clip]{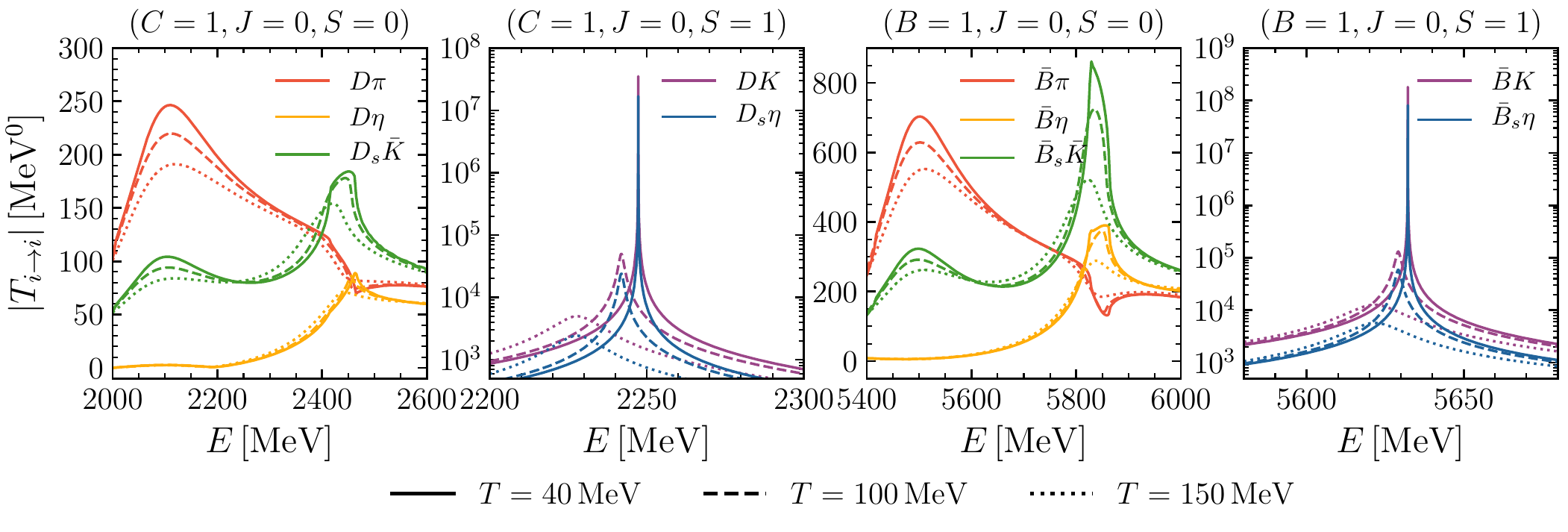}\vspace{-0.4cm}
\caption{Diagonal elements of the $T$-matrix in the sectors with charm $C=1$ (bottom $B=1$), spin $J=0$ and strangeness $S=0,1$. (Data partially taken from \cite{Montana:2020lfi}.)}
\label{fig:Tmatrix}       
\end{figure}

The states that are dynamically generated by our model also suffer a modification at finite temperature, as can be seen by looking at the unitarized scattering amplitudes plotted in Fig.~\ref{fig:Tmatrix} for $T=40,100,150$~MeV. In the sector with charm $C=1$, spin $J=0$ and strangeness $S=0$ the experimental $D_0^*(2300)$ is generated with a two-pole structure at $T=0$. These two resonances move towards lower energies and their widths increase with increasing temperatures. We note that the fitting of these structures has to be performed after background subtraction as explained in \cite{Montana:2020lfi}. 
In the strange sector, where the narrow $D_{s0}^*(2317)$ is experimentally seen, the bound state found in vacuum develops a width and its mass decreases in a thermal bath.
Figure~\ref{fig:Tmatrix} also shows the analogue thermal effects on the dynamically generated states in the bottom sector, where no experimental states have been reported yet.
\newpage


\vspace{-0.2cm}
\section{Open-charm Euclidean correlators with EFT}

Given the spectral function $\rho(\omega;T)$, the Euclidean correlator is calculated from the convolution with a temperature dependent kernel $K(\tau,\omega;T)$ as:
\begin{equation}\label{eq:corr}
    G_E(\tau;T)=\int_0^\infty d\omega\, K(\tau,\omega;T)\,\rho(\omega;T) \ , \;
     K(\tau,\omega;T)=\cosh\Bigg[\omega\Bigg(\tau-\frac{1}{2T}\Bigg)\Bigg]\sinh^{-1}\Bigg(\frac{\omega}{2T}\Bigg) \ .
\end{equation}

\begin{figure}[b!]\vspace{-0.2cm}
\centering
\includegraphics[width=9.4cm,clip]{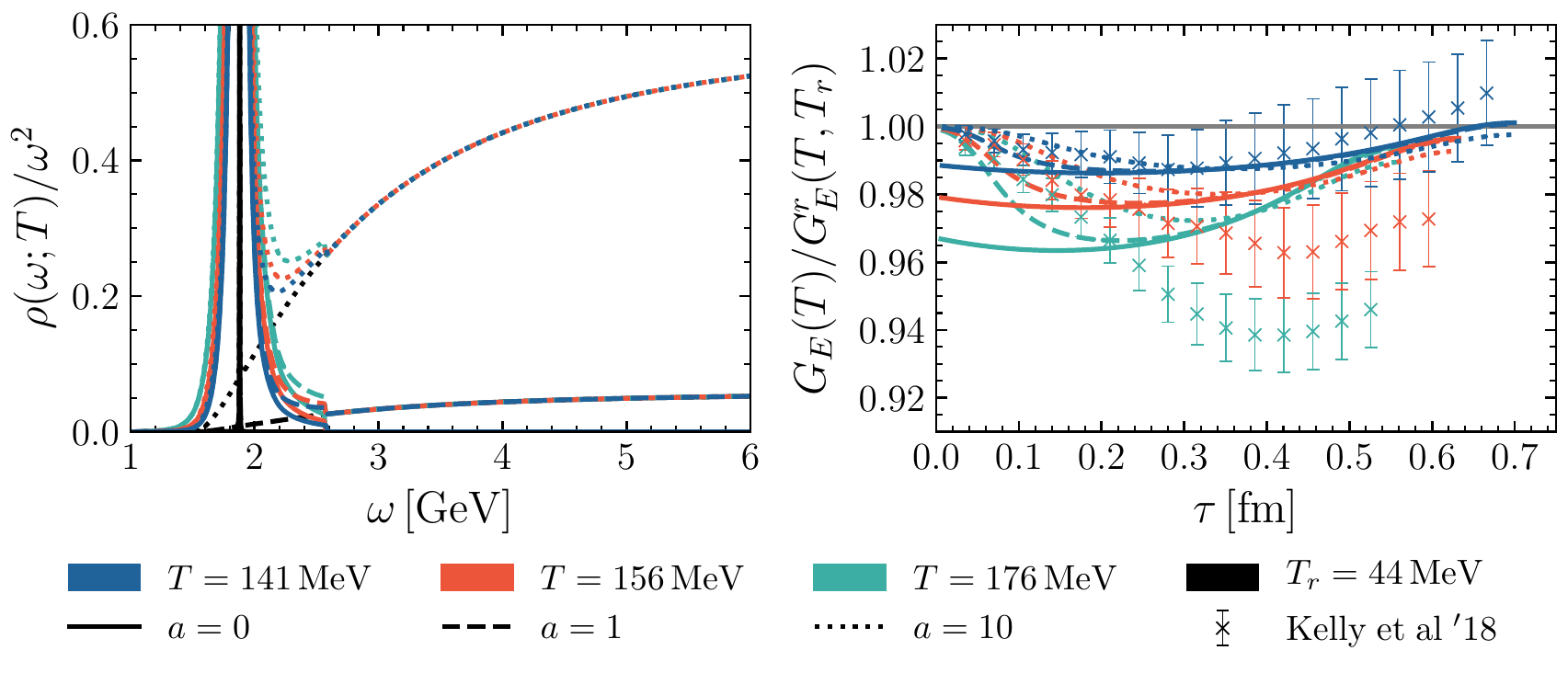}\vspace{-0.2cm}
\caption{Full spectral functions of the $D$-meson (left panel), and Euclidean correlators (right panel) for various temperatures below $T_c$ and values of the weight parameter $a$. (Taken from \cite{Montana:2020var}.)}
\label{fig:correlators}       
\end{figure}

Euclidean correlators are computed with LQCD and the inversion of Eq.~(\ref{eq:corr}) is required in order to extract the information about the physical states contained in the spectral function. This is an ill-posed problem and for this reason in \cite{Montana:2020var} we compare the thermal EFT with LQCD results directly at the level of the Euclidean correlators. We obtain the spectral function of the ground state using Eq.~(\ref{eq:S}), employing in the model the non-physical meson masses used in the LQCD calculations in \cite{Kelly:2018hsi}, and we define the full spectral function 
\begin{equation}\label{eq:fullS}
    \rho(\omega;T)= M_D^4S_D(\omega;T)+a\rho_{\rm cont}(\omega;T) \ ,
\end{equation}
where the second term corresponds to the continuum of scattering states and for which we take a parametrization of the free-meson spectral function in the non-interacting limit. The parameter $a$ weights the contribution of the continuum with respect to the ground-state.

The left panel of Fig.~\ref{fig:correlators} shows the full spectral function for different values of $a$ and temperatures below $T_c$, while the right panel shows the corresponding ratio of the Euclidean to the reconstructed correlators, compatible with LQCD within the errorbars for the lowest temperatures upon the inclusion of the continuum.

\section{Off-shell transport coefficients of $D$ mesons}
\begin{figure}[htbp!]
\centering
\sidecaption
\includegraphics[width=5.8cm,clip]{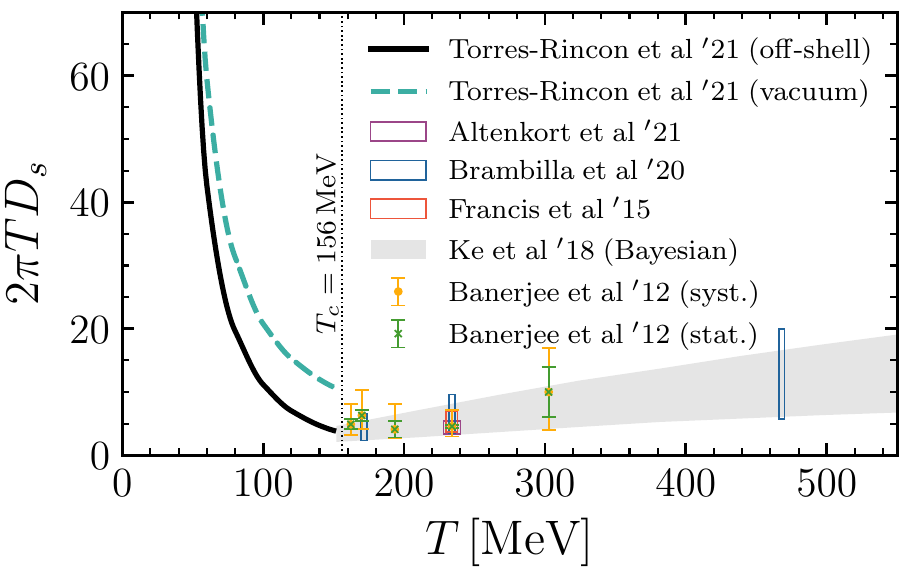}
\caption{Spatial diffusion coefficient below and above $T_c$ resulting from different approaches (see \cite{Torres-Rincon:2021yga} for details). The empty boxes correspond to a range of values of $2\pi TD_s$ for the central value of $T$. (Taken from \cite{Torres-Rincon:2021yga}.) }
\label{fig:transport}       
\end{figure}

In \cite{Torres-Rincon:2021yga} we apply the results of the EFT at non-zero temperature to the calculation of transport coefficients of an off-shell $D$-meson. We derive an off-shell Fokker-Planck equation for the Green's function,
\begin{equation}
    \frac{\partial}{\partial t} G_D^< (t,k) = \frac{\partial}{\partial k^i} \left\{ \hat{A} (k;T) k^i G_D^< (t,k) + \frac{\partial}{\partial k^j} \left[ \hat{B}_0(k;T) \Delta^{ij} + \hat{B}_1(k;T) \frac{k^i k^j}{|{\vec{k}|}^2} \right] G_D^< (t,k) \right\} \ ,
\end{equation}
in which the transport coefficients $\hat{A}$, $\hat{B}_0$ and $\hat{B}_1$ are defined off shell, as they are calculated using spectral functions, as well as thermal scattering amplitudes. The $D$-meson spatial diffusion coefficient $D_s(T)$, obtained from the static limit of the $\hat{B}_0$ at the quasiparticle energy, 
\begin{equation}
    2\pi TD_s(T)=\lim_{{\vec{k}}\rightarrow 0}\frac{2\pi T^{\,3}}{\hat{B}_0({E_k,\vec{k}};T)} \ ,
\end{equation}
is shown in Fig.~\ref{fig:transport} below $T_c$ in the case of the  full off-shell thermal calculation, together with the on-shell result using vacuum amplitudes. The main difference is the appearance of the so-called Landau contribution when using thermal scattering amplitudes, making it possible to smoothly match to recent results of LQCD calculations and a Bayesian analysis around $T_c$.

\vspace{-0.5cm}
\section*{Acknowledgements}
 G.M. acknowledges support from the FPU17/04910 Doctoral Grant from the Ministerio de Econom\'ia y Competitividad (MECD). G.M. and A.R. ackgnowledge support from the Ministerio de Econom\'ia y Competitividad (MINECO) under the project MDM-2014-0369 of ICCUB and, with additional European FEDER funds, under the contract FIS2017-87534-P. The research of L.T. has been supported by the Ministerio de Ciencia e Innovaci\'on and the European Regional Development Fund (ERDF) under contract PID2019-110165GB-I00 (Ref.10.13039/501100011033). L.T. and J.M.T.-R. acknowledge support from the Deutsche Forschungsgemeinschaft through projects no. 411563442 and no. 315477589 - TRR 211.

\bibliography{D-meson}

\end{document}